\documentclass[review]{elsarticle}

\usepackage{lineno,hyperref}
\modulolinenumbers[5]

\journal{Journal of \LaTeX\ Templates}

\biboptions{authoryear}

\begin{document}

\begin{frontmatter}

\title{A method to adjust the impedance of the transmission line in a Multi-Strip Multi-Gap Resistive Plate Counter}


\author[addr]{D.~Barto\c{s}}
\author[addr]{M.~Petri\c{s}}
\author[addr]{M.~Petrovici\corref{mycorrespondingauthor}}
\cortext[mycorrespondingauthor]{Corresponding author}
\ead{mpetro@nipne.ro}
\author[addr]{L.~R\u adulescu}
\author[addr]{V.~Simion}

\address[addr]{Hadron Physics Department,\\
National Institute for Physics and Nuclear Engineering, IFIN-HH\\
Bucharest, Romania\\}

\begin{abstract}
While in a triggered experiment the matching of the RPC transmission line impedance with the one of the front-end electronics is less critical, for a trigger-less data recording this becomes mandatory. As expected, impedance matching is not straightforward when other requirements in terms of time and position resolutions, efficiency and granularity, have to be fulfilled in the same time. 
A method and the very first results obtained with a RPC prototype built based on it, presented in this paper, show that the impedance matching, independent of its granularity, can be achieved using an innovative architecture of the RPC.     
\end{abstract}


\end{frontmatter}


\section{Introduction}

 The Compressed Baryonic Matter (CBM) experiment at the future Facility for Antiproton and Ion Research (FAIR) in Darmstadt is dedicated to the exploration of the QCD phase diagram at high net-baryon densities using high-intensity heavy-ion beams provided by the FAIR accelerators.
For particle identification, the experimental setup will include a Time of Flight (TOF) subsystem covering an active area of 120~m$^2$.
The CBM-TOF wall \cite{tdr} is based on Multi-Gap Resistive Plate Counters (MRPC) working in avalanche mode \cite{mrpc},  with strip structured readout electrodes - Multi-Strip Multi-Gap Resistive Plate Counters (MSMGRPC) \cite{pet1} read-out in a differential mode \cite{jinst2012}. 

The system time resolution, including the contribution of the reference detector and associated electronics, has to be better than 80~ps, with an efficiency better than 95\%. For 10~MHz minimum bias Au+Au collisions, the innermost part of the detector is exposed to counting rates up to 25~kHz/cm$^2$ and high particle multiplicities.

The granularity of the TOF wall is required to correspond to an occupancy bellow 5\% in the most inner zone of the CBM-TOF wall. 
Under these conditions, the effective area of a single read-out cell has to be $\sim$6~cm$^2$ in the central part of the wall, close to the beam pipe. However, since the track density drops rapidly with increasing distance from the beam axis, the effective cell size at the increased polar angles could be larger, as it is shown in reference~\cite{tdr}. 
As a function of polar angle, for a given strip pitch, the length of the readout - strips can be easily adjusted to the required granularity. 

In high interaction rate experiments, as CBM is designed to be used, a trigger-less readout concept is required. In a continuous readout operation all signals passing the electronics thresholds are digitised, time stamped and processed.
This imposes to the  MSMGRPCs, a perfect matching of the impedance of the signal transmission line to the input impedance of the front-end electronics, in order to reduce the large amount of fake information resulted from reflexions.

The transmission line corresponding to a narrow strip of a RPC prototype (2.54 mm pitch with 1.1 mm strip width) developed in our group \cite{jinst2012} has characteristic impedance of 100~$\Omega$, matching the input impedance of the differential front-end electronics used for signal processing \cite{nino}. 
Besides a very good time resolution, in the region of 50 ps, and efficiency better than 95$\%$,
such an architecture gives access to a two-dimensional position information with a resolution of 400 $\mu$m across the strips and $\sim$4.5 mm along the strip direction. 
The number of electronics channels required to equip the most forward polar angles of the CBM~-~TOF wall with such type of MSMGRPC of $\sim$140,000 has a direct impact on the final cost of the whole subdetector. 
This number is considerably reduced using a 7.4 mm strip pitch. A prototype 
with a 7.4~mm strip pitch (5.6~mm width) has been built and tested \cite{JoP2016}.   
This prototype, based on low resistivity glass \cite{Yiwang}, demonstrated excellent performance in terms of time resolution and efficiency up to local particle flux of 10$^5$~particles/(cm$^2$$\cdot$s) and up to an exposure of 10$^4$~particles/(cm$^2$$\cdot$s) all over the counter surface.
Due to the larger strip width, a symmetric 2x5 gas gaps structure and properties of the low resistivity glass, the differential transmission line defined by the corresponding strips of the readout electrodes has a 50~$\Omega$ impedance. Therefore, an impedance matching with the input impedance of fast amplifiers of 100~$\Omega$ was required at the level of FEE motherboards. 
With a proper choice of the strip length, the readout architecture of this prototype fulfils the required granularity for the most inner zone of the CBM-TOF wall with $\sim$3 times less electronic channels relative to the narrow strip solution mentioned above. 

In order to fulfil simultaneously all requirements in terms of detector performance, granularity and impedance matching, an original solution is proposed. We are reporting here the design of a new MSMGRPC prototype with a strip pitch of 7.2 mm, which matches any input impedance of a given front-end electronics by tuning the impedance of the signal transmission line through the value of the readout strip width, independent on the granularity given by the width of the high voltage strip and cluster size. The results of very preliminary measurements which confirm the expectations based on simulations are presented. 

\section{Detector inner architecture}

\begin{figure}[h!tb]
\centering
\includegraphics[width=0.95\textwidth]{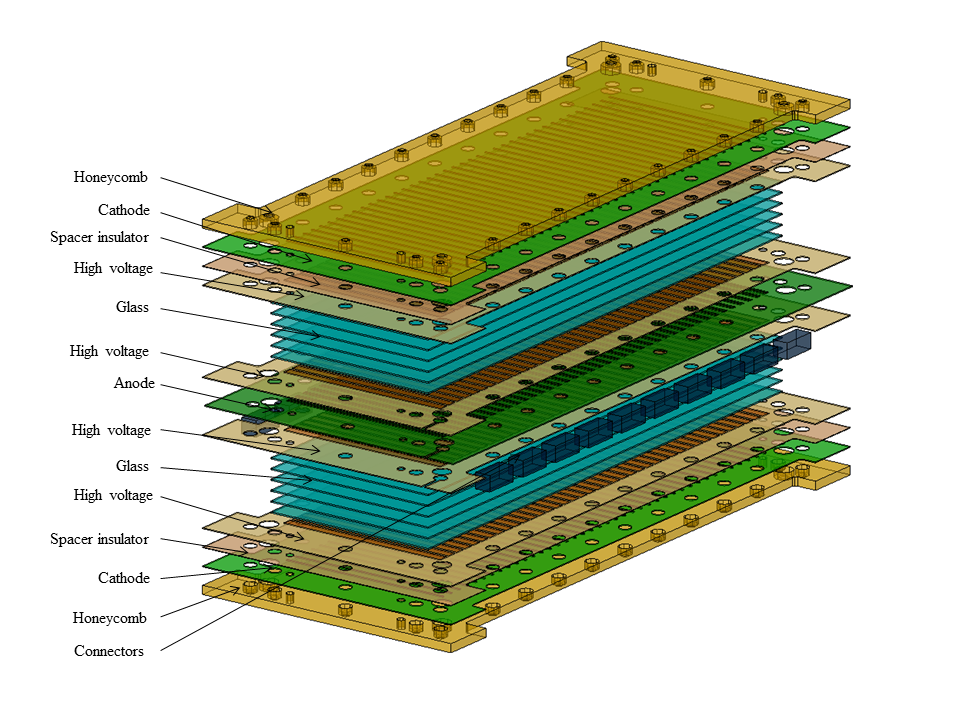}
\caption{3D exploded view of the detector structure.}
\label{fig1}
\end{figure}

The detector inner geometry, schematically presented in Fig.~\ref{fig1}, is the same as for the previous prototypes already reported \cite{jinst2012, jinst2016}. It has a structure of two stacks symmetrically disposed relative to the central readout electrode. Each stack contains six plan parallel resistive electrodes of 0.7~mm thickness, equally spaced by five gas gaps. The size of the gas gap is defined by the 140~$\mu$m diameter nylon fishing line used as spacer. The resistive electrodes are made from low resistivity  glass ($\sim$1.5x10$^{10}$~$\Omega$cm). 

The outermost glass plates of each stack are in contact with the Cu strips of the cathode electrodes made of FR4 material of 0.5~mm thickness. An insulator of 300 $\mu$m thickness prevent discharges from the high voltage (HV) electrodes to the readout electrodes. 
The individual HV strips are separated from the common HV strip by 12~k$\Omega$ resistors in order to decrease the influence of the electric field distortion due to the avalanche developed in the region of a given strip on the neighbouring ones.
 
\begin{figure}[h!tb]
\centering
\includegraphics[width=0.75\textwidth]{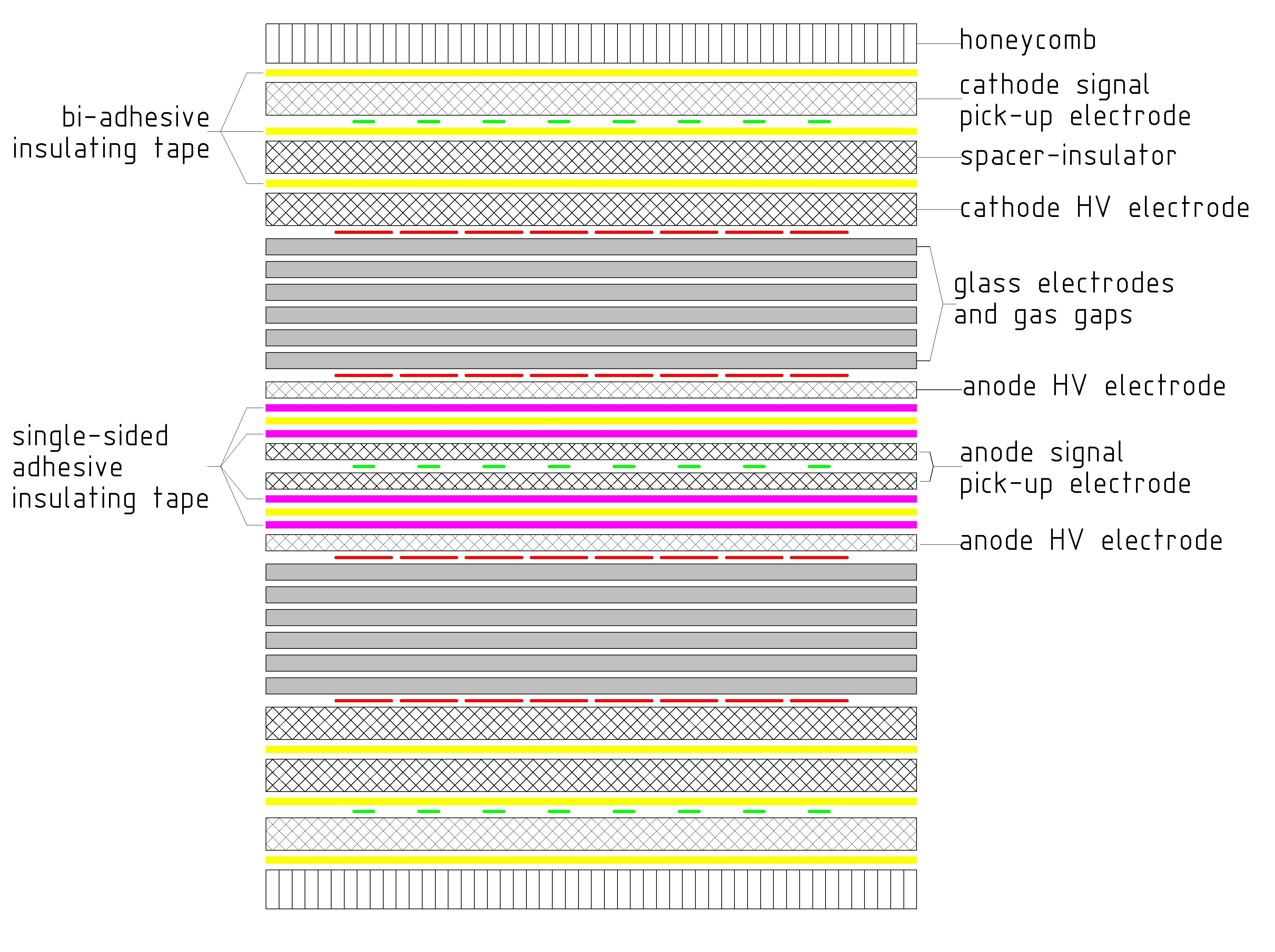}
\caption{A cross-section through the detector across the strips.}
\label{fig2}
\end{figure}

The central read-out electrode is a single layer Cu strip structure, sandwiched between two thin layers of FR4 of 0.25~mm thickness each. The two anode electrodes, sandwiching the central read-out electrode are identical with the cathode ones. 
The HV electrodes define an active area of 300~mm~x~96~mm.
The electric field created in this way between the anode and cathode corresponding HV strips improves the time resolution in high counting rate environment by removing faster the space charge from the active gas volume.
The mechanical stability and precise alignment of the electrodes is maintained by two honeycomb plates positioned on the outer sides of the two stacks, as it is shown in the cross section across the strips presented in Fig.~\ref{fig2}. 
The signals are readout in a differential mode, both, the anode and the cathode signals, being fed into the input of a readout electronics channel. The readout strips behind the corresponding anode and cathode HV ones define a signal transmission line of which impedance depends on the strip width and the properties of the whole structure in between. 

In order to 
 achieve a direct matching with the 100~$\Omega$ input impedance of the front-end electronics and the required granularity, 
an innovative solution is proposed: for a given strip pitch of the HV electrodes, which is essential for the detector granularity, one could vary the signal strip width such to obtain a desired impedance of the transmission line, 100 $\Omega$ in our case. 

\section{Simulation of the transmission line impedance} 
\label{sec3}
A transmission line is a pair of parallel conductors exhibiting certain characteristics due to distributed capacitance and inductance along its length. The characteristic impedance of a transmission line is equal to the square root of the ratio of the line inductance per unit length {$\displaystyle L$} divided by the line capacitance per unit length {$\displaystyle C$}, 
\begin{equation}
 Z_0 = \sqrt{L/C} ; \hspace{0.3cm} where \hspace{0.3cm} C = \epsilon_0\epsilon_r(w/h);
\label{eq1}
\end{equation}
Through the capacitance per unit length the characteristic impedance depends on the distance between the two conductors ($\displaystyle h$), the width of metalic layer ($\displaystyle w$) and the relative permitivity of the insulator between them ($\epsilon_r$). One can derive from Eq.~\ref{eq1} that the transmission lineâs characteristic impedance ($\displaystyle Z_0$) increases as the conductor spacing $\displaystyle h$ increases and the width $\displaystyle w$ of methalic layer of the line decreases. 

The strip width of the readout electrodes  was decided based on the results of signal propagation simulation using APLAC \cite{aplac} for the RPC architecture presented in the previous chapter.
APLAC is a commercial software (high frequency simulation technology), commonly used for such simulations.  It was used also by us in estimating the  transmission line impedance for the previous MSMGRPC prototypes and the predictions were confirmed by the measurements. 

As input for the simulation, the transmission line for the signal propagation was defined as a multilayer structure composed from the cathode and anode readout strips separated by the PCB layers, kapton foils, resistive glass electrodes and gas layers for one single stack of the counter (see figure~\ref{fig1}). 
\begin{figure}[h!tb]
\centering
\includegraphics[width=0.55\textwidth]{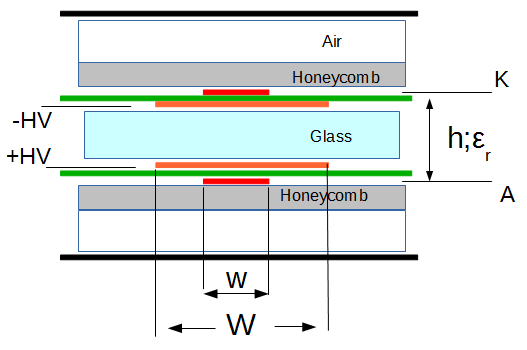}
\caption{Equivalent detector structure considered in the simulations.}
\label{fig3}
\end{figure}
The individual dielectric layers positioned between the anode and cathode readout strips were considered as capacitors coupled in series.
The values of the thickness and permittivity of the individual layers (glass, kapton, FR4, gas) were 
considered in the calculation as an equivalent dielectric thickness {$\displaystyle h$} and equivalent 
dielectric constant, $\epsilon_r$, as it is shown in figure~\ref{fig3}.
 A differential signal with  timing  characteristics of 50 ps rise time, 300 ps width and 50 ps fall time  was 
injected at the input of the transmission line. The opposite end of the transmission line was connected differentially to a load resistor with 
{$\displaystyle Z_L$} impedance . 
The input/output signals were recorded for different values of the readout strip width {$\displaystyle w$}.
\begin{figure}[h!tb]
\centering
\includegraphics[width=0.75\textwidth]{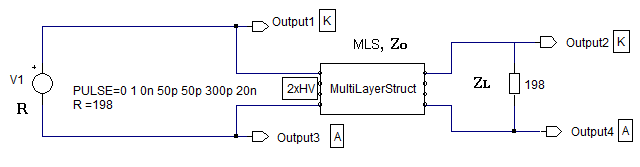}
\caption{Simulation scheme.}
\label{fig4}
\end{figure}

\begin{figure}[htb]
\centering
\vspace{-0.20cm}
\includegraphics*[width=0.6\textwidth]{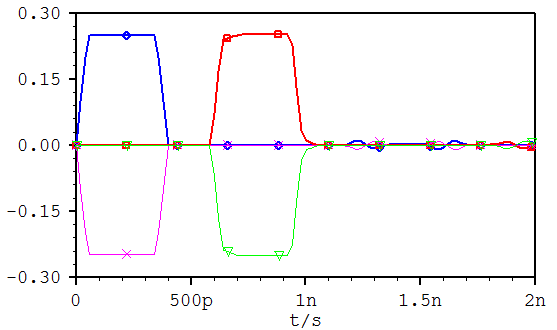}
\vspace{-0.35cm}
\caption{Simulated signals picked - up on the anode (input - magenta, output - green) and cathode (input - blue, output - red) electrodes.}
\label{fig5}
\end{figure}
If the transmission line is matched to the input and output impedances at the two ends, i.e.:
\begin{equation}
R=Z_0=Z_L 
\end{equation}
no loss in the transmission line should be observed. 
 The simulated signals picked - up from the anode (input - magenta, output - green) and cathode (input - blue, output - red) electrodes for one half of the structure are shown in Fig.~\ref{fig4}. They are injected by a pulse generator with 198~$\Omega$ internal resistor on the read-out strip and read-out on a 198~$\Omega$ load resistor on the other side. 
  The output signals reproduce very well the input ones, without any visible distortions of their shape and magnitude, showing that the transmission line is matched to the input/output impedances.
The delay of the output signals relative to the input ones is of {$\displaystyle \Delta t$}~=~0.65~ns for {$\displaystyle l$}~=~9.6~cm strip length. This implies in a propagation signal velocity {$\displaystyle v_s$} on the transmission line of:
\begin{equation}
v_s = \frac{l}{\Delta t} = 14.76~cm/ns
\end{equation}

Therefore, for a 1.32~mm strip width, 0.7~mm glass thickness, 140~$\mu$m gas gap, and relative glass permitivity $\epsilon_r$~=~9.1,  APLAC simulations predicted a transmission line impedance of {$\displaystyle Z_0$} = 198~$\Omega$. As it was specified above, this value was estimated for a single stack, representing one half of the structure. 
\begin{figure}[htb]
\centering
\vspace{-0.20cm}
\includegraphics*[width=0.95\textwidth]{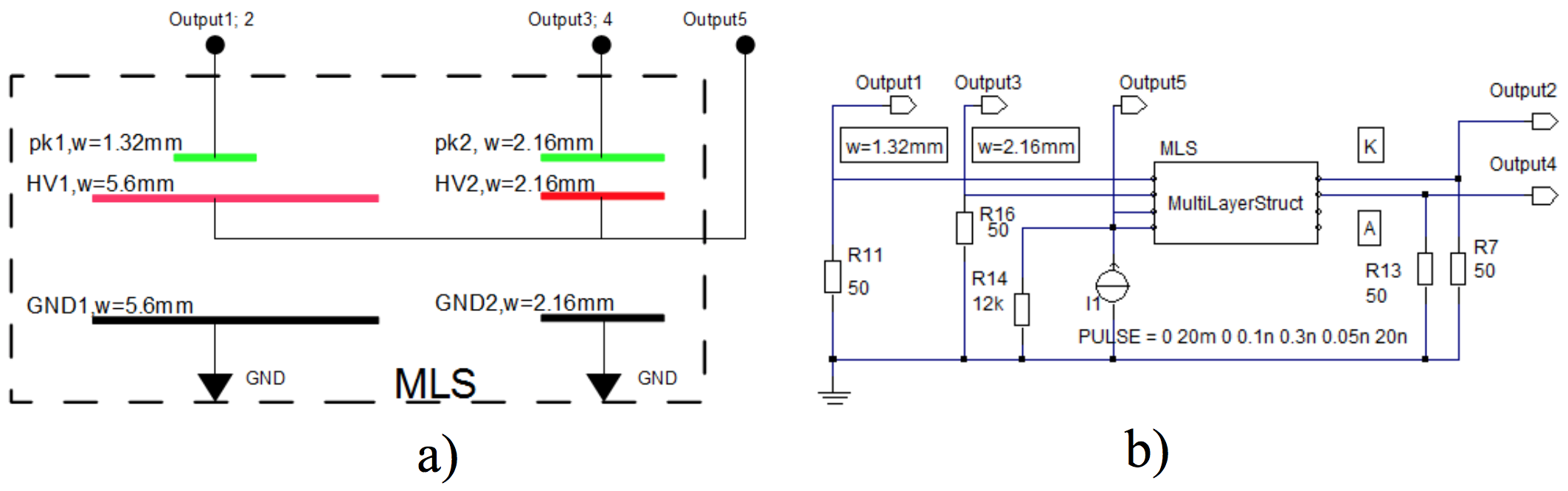}
\caption{Simulation scheme used to study the amplitude of the induced signal on a strip of the pick-up electrode. a) left side corresponds to the architecture described in the preset section, the width of the strip signal being smaller than the width of the HV electrode while the right side corresponds to a narrow strip configuration where both, the signal and HV strips have the same width; b) the equivalent scheme used in the APLAC simulation}
\label{fig6}
\end{figure} 
As the corresponding transmission lines of the two stacks are connected in parallel, the equivalent impedance of a MSMGRPC transmission line is {$\displaystyle Z0$}/2~=~99~$\Omega$. 
This value is matched to the input impedance of the front-end electronics (FEE) used for MSMGRPC signal processing, as it is based on NINO chip developed within ALICE-TOF collaboration \cite{nino} or PADI developed within the CBM-TOF collaboration \cite{padi}. 
In order to estimate the consequence of such an architecture on the amplitude of the signal induced on the pick-up electrodes relative to the case when the signal strip have the same width as the HV ones, we used in the APLAC simulation the scheme presented in Fig.~\ref{fig6}. Fig.~\ref{fig6}a left side corresponds to the architecture presented in this Section and the 
narrow strip configuration where both, the signal and HV strips have the same width
is presented in Fig.~\ref{fig6}a right side. The equivalent scheme used for APLAC simulation is presented in Fig.~\ref{fig6}b.
The results are presented in Fig.~\ref{fig7}. The injected signal on the HV strip is represented by black line. The red and green pulses correspond to the induced signals on the near and far end side of the corresponding pick-up signal strip with the same width as the HV one. The blue and dark green pulses correspond to the induced signals on the near and far side of the corresponding pick-up signal strip narrower than the HV one. For the second configuration, a reduction of $\sim$6\% in the amplitude of the signal induced on the near side is observed. The difference in the amplitude at the far side signal corresponding to the two architectures is bellow 2\%. Therefore, no deterioration in the detector efficiency is expected. 
\begin{figure}[htb]
\centering
\vspace{-0.20cm}
\includegraphics*[width=0.85\textwidth]{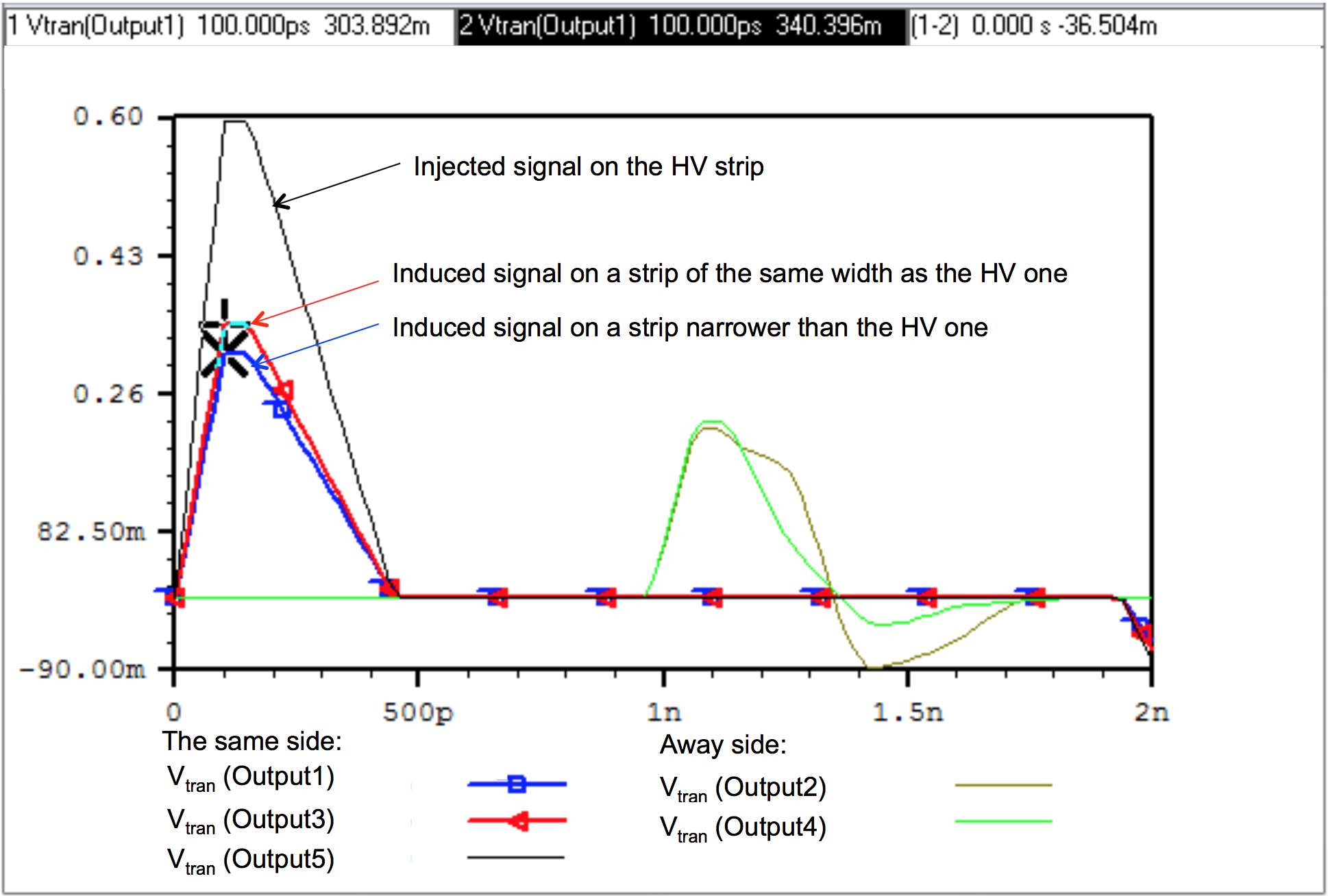}
\caption{Result of APLAC simulation using the scheme presented in Fig.~\ref{fig6}. Black line - the signal injected at one end of the HV strip; red and green pulses correspond to the induced signals on the near and far side of the corresponding pick-up signal strip of the same width as the HV one - right side configuration in Fig.~\ref{fig6}a; blue and dark green pulses correspond to the induced signals on the near and far side of the corresponding pick-up signal strip with narrower width relative to the HV one - left side configuration in  Fig.~\ref{fig6}a}
\label{fig7}
\end{figure} 

\section{Detector prototype}

 Based on the results of APLAC simulations presented in the previous chapter, was designed and constructed a MSMGRPC prototype following the general structure of previous ones, presented in Fig.~\ref{fig1} with HV strip width of 5.6 mm, 1.32 mm width for the strips of the corresponding read-out electrodes and the pitch size of 7.2 mm for all electrodes. The expected cluster size is in the range of $\sim$ 2 strips which together with a proper strip length leads to the required granularity of the innermost zone of the CBM-TOF detector.
\begin{figure}[htb]
\centering
\vspace{-0.20cm}
\includegraphics*[width=0.95\textwidth]{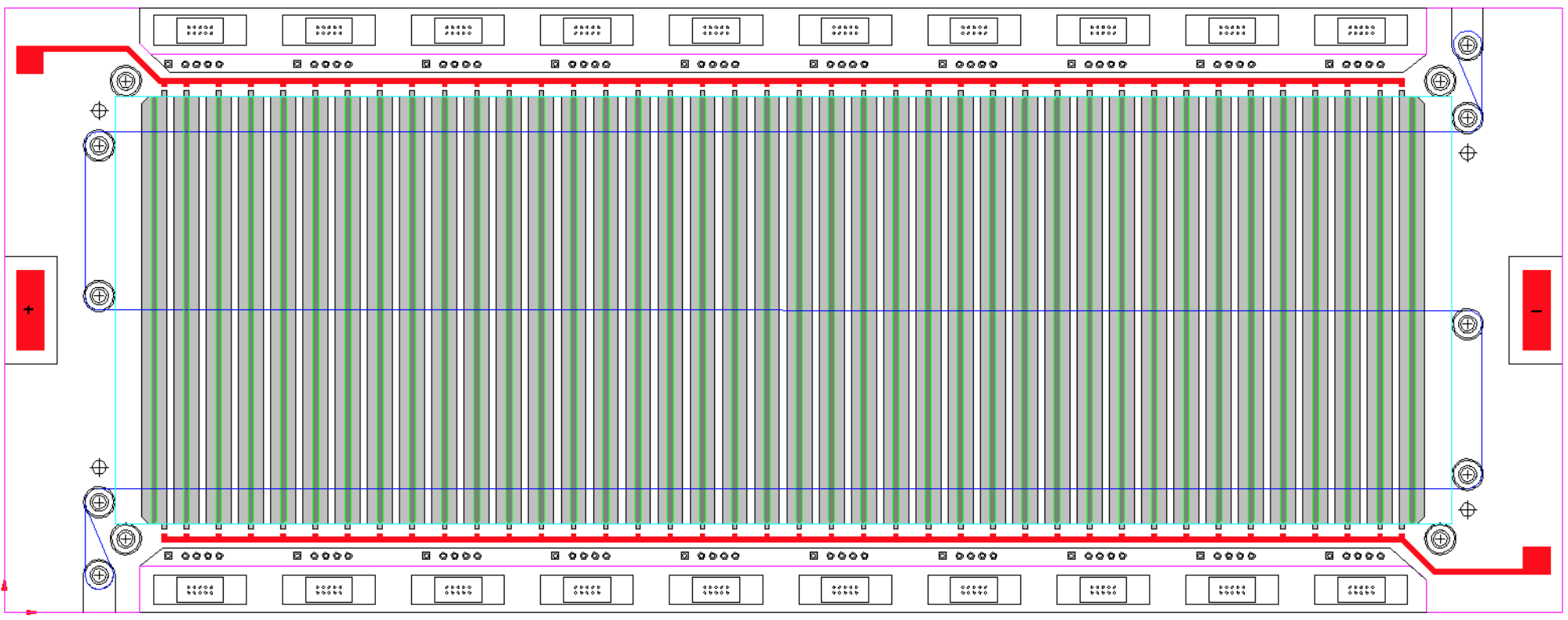}
\caption{Schematic front-view cross section of the detector.}
\label{fig8}
\end{figure} 
 The narrow readout strips are centered on the wider HV strips. Details on the spacer routing between the resistive electrodes, high voltage distribution and position of the signal connectors on the edges of the pick-up signals electrodes can be followed in the front-view cross section presented in Fig.~\ref{fig6}. 
Due to the particular strip structure of the HV electrodes positioned under the readout electrode, the signal is first induced on the HV electrodes and subsequently on the readout electrodes. 
 A schematic view of a single readout channel, including the MSMGRPC transmission line, is depicted in Fig.~\ref{fig9}.
\begin{figure}[htb]
\centering
\includegraphics*[width=125mm]{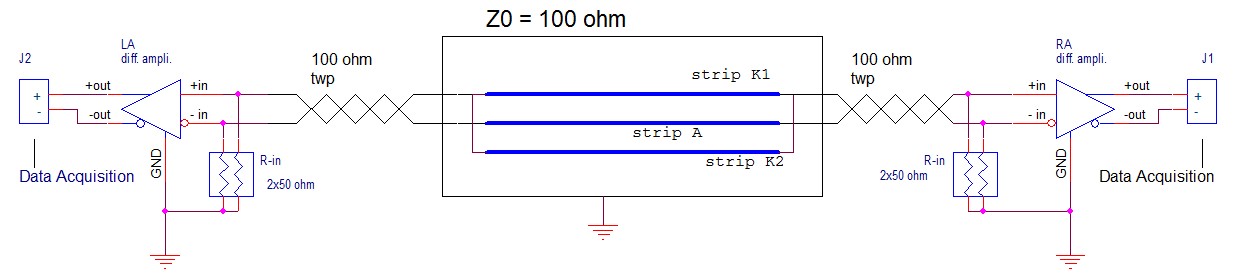}
\caption{A schematic view of a single readout channel.}
\label{fig9}
\end{figure}
   A photo of the assembled stack with the structure presented in Fig.~\ref{fig1} is shown in Fig.~\ref{fig10} left side. In the same period it was assembled a single stack MSMGRPC structure with 8 gas gaps and 100 $\Omega$ impedance. The two RPCs were assembled mechanically on top of each other for cosmic ray and in-beam tests. The final structure with the cabled signals can be followed in Fig.~\ref{fig10} right side. 
 The signal transmission to the FEE plugged into the connectors mounted on the back panel of the tight gas box housing the detector is made by twisted pair cables with 100~$\Omega$ characteristic impedance, see Fig.~\ref{fig8}. 
  
 The time of flight information, independent on the position of the induced signal along the strip is based on the mean value of the times measured at the two ends, while the position along the strip is obtained from the difference of the two time signals. Charge sharing among the strips gives the position information across the strips.   

\section{Detector laboratory test}

\begin{figure}[htb]
\centering
\vspace{-0.20cm}
\includegraphics*[width=1.0\textwidth]{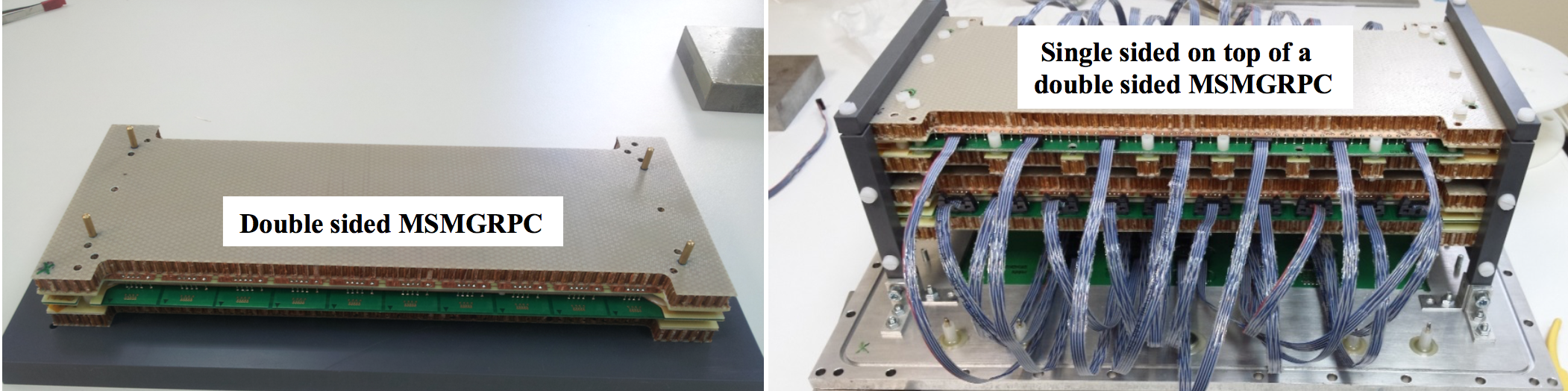}
\caption{RPC prototype mounted on an Aluminium backplane on which the special PCBs with the feed through signal connectors on.}
\label{fig10}
\end{figure}

\begin{figure}[htb]
\centering
\includegraphics*[width=0.60\textwidth]{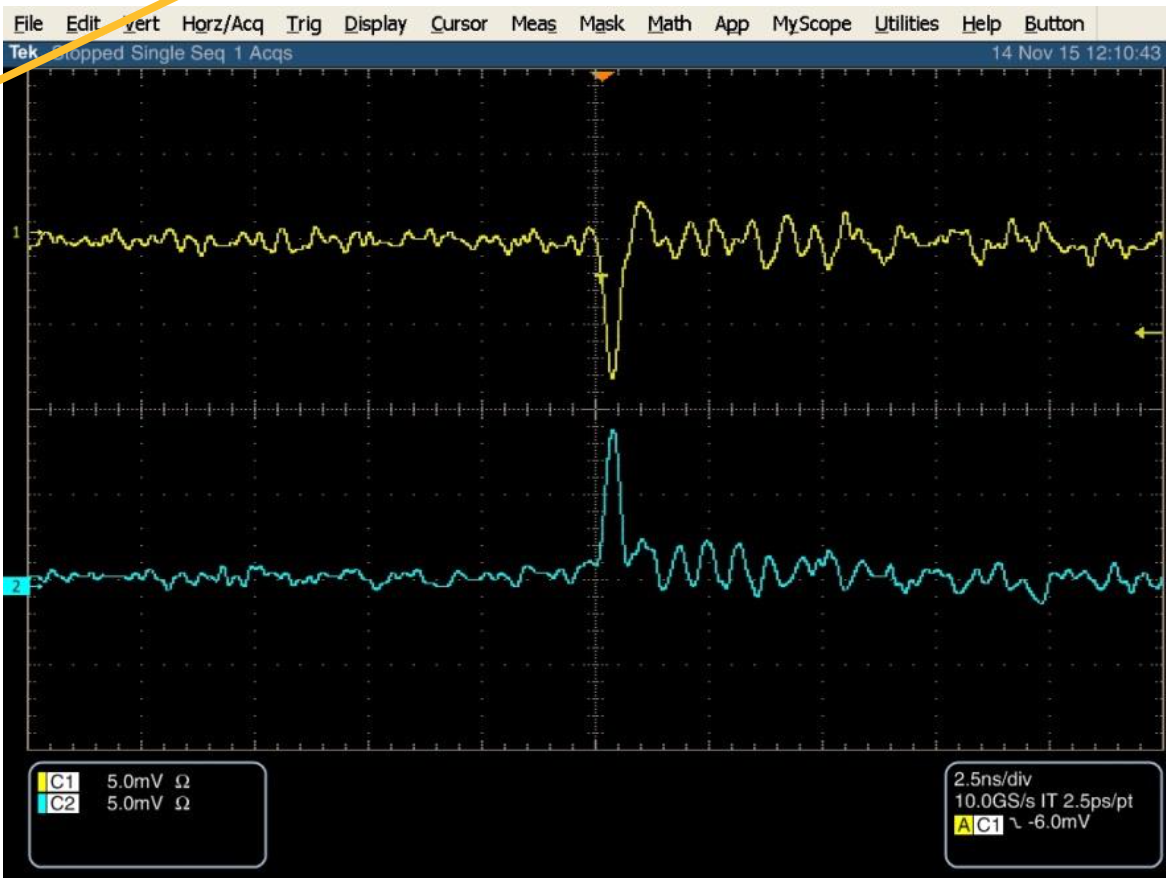}
\caption{Direct anode (yellow) and cathode (blue) pick-up signals from one side of a given strip produced by cosmic rays.}
\label{fig11}
\end{figure}

The prototype with the characteristics described in the previous chapter and presented in Fig.~\ref{fig10} 
was tested with a gas mixture of 90\%C$_2$H$_2$F$_4$+10\%SF$_6$ and electric field of 157 kV/cm. An example of typical signals produced by cosmic rays is shown in Fig.~\ref{fig11}.

\begin{figure}[htb]
\centering
\includegraphics*[width=0.60\textwidth]{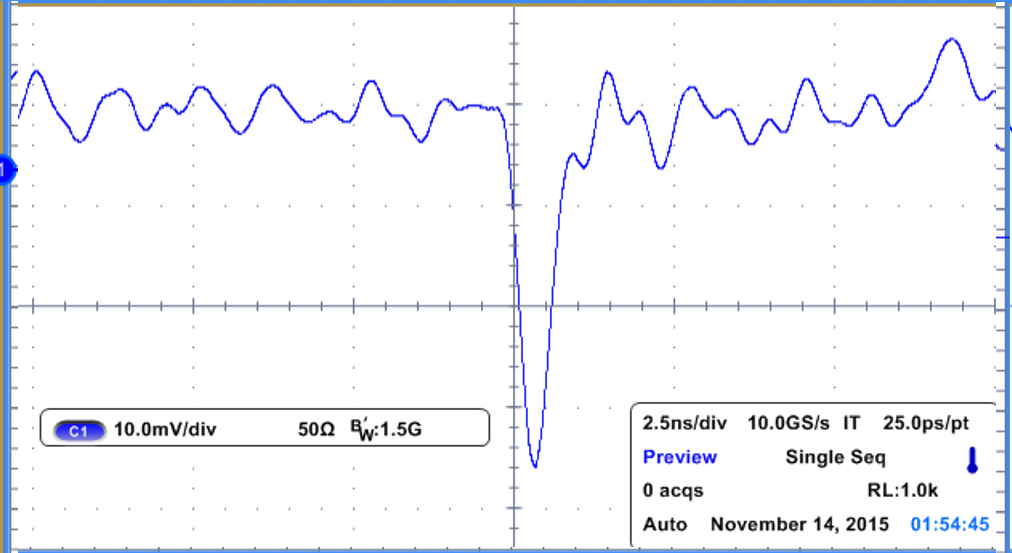}
\caption{Direct pick-up signals produced by cosmic rays recorded with a differential probe.}
\label{fig12}
\end{figure}

They were recorded using a Tektronix TDS 7254B 3~GHz oscilloscope  connected directly to signal connectors of the back flange corresponding to the ends of the detector transmission lines of the pick-up electrodes behind the anode and cathode HV electrodes, respectively, on one side. The corresponding opposite ends were terminated by 50 $\Omega$. Fast signals with rise time of few hundred picoseconds and  $<$1~ns width, corresponding to that specific strips, without any reflections on the displayed time scale of 7.5~ns are evidenced. 
A sample of direct pick-up signals produced by cosmic rays recorded with a differential probe
can be followed in Fig.~\ref{fig12}.
   As could be observed, none of the read-out modes show any reflected signals expected to appear in case of unmatched impedances at the level of the connectors, on the RPC PCBs or after the 30 cm twisted cables on the back flange of the housing box. 

\section{Conclusions}

In the present paper is presented a method to tune the MSMGRPC impedance such to match the input impedance of the corresponding FEE. This can be achieved independent of the RPC granularity by using an innovative RPC architecture. The very first results obtained with an RPC prototype built based on this method support the expectations based on APLAC simulations.  

{\bf Acknowledgments}

This work was carried out within the PN 16 42 01 04 and F04 HICORDEFEND supported by Ministry of Research and Inovation and IFA coordinating agency. 

{\bf References}

\bibliographystyle{unsrt}


\end{document}